\documentclass{elsart}
\usepackage{graphicx}
\usepackage{amsmath}
\usepackage{amsfonts}
\usepackage{amssymb}

\begin{document}
\begin{frontmatter}

\title{Hamilton-Jacobi equation and the breaking of the WKB approximation}
\author{F. Canfora}
\address{Universit\`{a} di
Salerno, Dipartimento di Fisica ''E.R.Caianiello'', Istituto
Nazionale di Fisica Nucleare, GC di Salerno, Via S.Allende, 84081
Baronissi (Salerno) Italy; e-mail: canfora@sa.infn.it, Phone:
+39-089-965228, Fax: +39-089-965275.}

\begin{abstract}
A simple method to deal with four dimensional Hamilton-Jacobi
equation for null hypersurfaces is introduced. This method allows to
find simple geometrical
 conditions which give rise to the failure of the WKB approximation on curved spacetimes.
The relation between such failure, extreme blackholes and the Cosmic
Censor hypothesis is briefly discussed.
\end{abstract}

\end{frontmatter}

\section{Introduction}

\noindent Hamilton-Jacobi (henceforth, HJ) equation is one of the most deeply
analyzed equation of theoretical physics because of its application in many
fields of great interest such as theoretical optics, quantum mechanics,
Hamiltonian dynamics. In gravitational physics many important applications are
based on HJ equation as, for instance, gravitational lensing. HJ equation also
encodes the light-cones structure of space-time and its analysis is also
fruitful to disclose the asymptotic symmetries of space-time itself. In
blackholes physics also, since the Carter's works \cite{Ca68a} \cite{Ca68b},
HJ equation played a prominent role in clarifying many important local as well
as global properties of such interesting objects. A seemingly new method to
deal with the curved four dimensional HJ equation seems to be able to disclose
interesting relations between complex structures, BPS states in gravity and
the breaking of the WKB approximation on curved space-time. In particular,
this method discloses in a clear and simple way the relation between the
impossibility to adopt the geometric optics approximation on certain curved
backgrounds and the presence of naked singularities.

\section{The method}

In the limit of small wave-lenghts, beams of null geodesics describing, for
example, electromagnetic or gravitational waves can be analyzed through the
WKB approximation. The rapidly oscillating factor multiplying the slowly
varying amplitude of the wave is usually written as $\exp(-iS)$\ where $S$ is
the eikonal function representing the wave fronts.\ On a four-dimensional
curved manifold $(M,g)$, $S$ satisfies the following HJ equation:
\begin{equation}
g^{\mu\nu}\partial_{\mu}S\partial_{\nu}S=0 \label{1}%
\end{equation}
$g^{\mu\nu}$ being the inverse metric. To exploit the quaternionic structure
of four dimensional light cones it is convenient to use a non holonomic base:
\begin{align}
g  &  =\eta_{ab}\theta^{a}\theta^{b}\nonumber\\
\theta^{a}  &  =\varepsilon_{\mu}^{(a)}dx^{\mu},g_{\mu\nu}=\eta_{ab}%
\varepsilon_{\mu}^{(a)}\varepsilon_{\nu}^{(b)}\label{L}\\
d\theta^{a}  &  \neq0\nonumber
\end{align}
where the greek indices $\mu$ run over space-time coordinates, the latin
indices $a=0,..,3$\ run over the tetrad basis $\varepsilon_{\mu}^{(a)}$ and
$\eta_{ab}$\ is the Minkowski tensor on tetrad indices. We will denote with
$\varepsilon_{(a)}^{\mu}$ the inverse tetrad. In terms of the tetrad
components eq. (\ref{1}) reads:
\begin{equation}
\left(  \varepsilon_{(0)}^{\mu}\partial_{\mu}S\right)  ^{2}=\overset
{3}{\underset{i=1}{\sum}}\left(  \varepsilon_{(i)}^{\mu}\partial_{\mu
}S\right)  ^{2}\text{.} \label{2}%
\end{equation}
Equation (\ref{2}) can be rewritten in the following useful form:
\begin{equation}
\det\Theta=0,\ \Theta=\left\{
\begin{array}
[c]{cc}%
\varepsilon_{(0)}^{\mu}\partial_{\mu}S+\varepsilon_{(1)}^{\mu}\partial_{\mu
}S & \varepsilon_{(2)}^{\mu}\partial_{\mu}S+\sqrt{-1}\varepsilon_{(3)}^{\mu
}\partial_{\mu}S\\
\varepsilon_{(2)}^{\mu}\partial_{\mu}S-\sqrt{-1}\varepsilon_{(3)}^{\mu
}\partial_{\mu}S & \varepsilon_{(0)}^{\mu}\partial_{\mu}S-\varepsilon
_{(1)}^{\mu}\partial_{\mu}S
\end{array}
\right\}  . \label{3}%
\end{equation}
It is well known that the determinant of a matrix is zero if and only if
either two rows (two columns) are proportional or one row (one column) vanishes.

Let us firstly consider the case in which $S$ is a real function. When the
first row vanishes one obtains:
\begin{align}
\varepsilon_{(0)}^{\mu}\partial_{\mu}S+\varepsilon_{(1)}^{\mu}\partial_{\mu}S
&  =0\label{11}\\
\varepsilon_{(2)}^{\mu}\partial_{\mu}S+\sqrt{-1}\varepsilon_{(3)}^{\mu
}\partial_{\mu}S  &  =0\text{.} \label{12}%
\end{align}
Separating the real and the imaginary part in eq. (\ref{12}) and rewriting eq.
(\ref{11}) we get
\begin{align}
\varepsilon_{(2)}^{\mu}\partial_{\mu}S  &  =0\label{P1}\\
\varepsilon_{(3)}^{\mu}\partial_{\mu}S  &  =0\label{P2}\\
\left(  \varepsilon_{(0)}^{\mu}+\varepsilon_{(1)}^{\mu}\right)  \partial_{\mu
}S  &  =0\text{.} \label{14}%
\end{align}
The above simple system of equations for $S$ tells us that locally $S$ has a
non trivial Lie derivative only along the tetrad $\varepsilon_{(0)}^{\mu
}-\varepsilon_{(1)}^{\mu}$ so that $S$ can be easily found by solving eq.
(\ref{14}) through the method of characteristic. Eqs. (\ref{14}), (\ref{P1})
and (\ref{P2}) could be the starting point of the WKB analysis in
gravitational lensing, but usually one only considers the characteristics of
the above equations, that is, usually only the light rays analysis is
performed (see, for example, \cite{SEF92}). It is clear that $S$ carries more
physical informations then single light rays, so, as far as gravitational
lensing is concerned, this method to treat HJ equation could be useful to
obtain global properties of the light beams such as the shear, the curvature
of the wavefronts and so on.

When the two rows of $\Theta$\ are proportional (the function of
proportionality $k$ being an arbitrary non vanishing function on $M$) and $S$
is real, the equations for $S$ can be written as follows:
\begin{align}
\left(  k-\frac{1}{k}\right)  \varepsilon_{(0)}^{\mu}\partial_{\mu}S  &
=\left(  k+\frac{1}{k}\right)  \varepsilon_{(1)}^{\mu}\partial_{\mu
}S\label{RE1}\\
2\varepsilon_{(0)}^{\mu}\partial_{\mu}S  &  =\left(  k+\frac{1}{k}\right)
\varepsilon_{(2)}^{\mu}\partial_{\mu}S\label{RE2}\\
\varepsilon_{(3)}^{\mu}\partial_{\mu}S  &  =0\text{.} \label{RE3}%
\end{align}
The above equations, with a suitable choice of $k$, can describe interesting
phenomena such as geodesics wrapping round and round in the $(2)-(3)$
direction. It could be rather fruitful to exploit the above method together
with an interesting method developed in \cite{Fri99a} \cite{Fri99b} to deal
with the singularities of light cones which can be described in a rather
detailed way. Indeed, one of the most important topic in which the above
scheme could be an important tools is holography (see, for two detailed
reviews, \cite{Bo02} \cite{AG00}). In the elegant framework of covariant
entropy bounds \cite{Bo99} \cite{Bo99b} which refined pioneering ideas of
Bekenstein \cite{Be81} 't Hooft \cite{tH93} and Susskind \cite{Su95}, an
important role to deduce the bounds on the entropy of a given two dimensional
spacelike surface $\Sigma$ is played by the null hypersurface spanned by the
non expanding light rays emanating from $\Sigma$ which, eventually, terminate
at caustics. The analysis of such null hypersurfaces is mainly based on
geodetics and Raychaudhuri equations. Any null hypersurfaces can be seen as a
solution of eq. (\ref{1}) so it is clear that the method here proposed could
be very important in a holographic perspective: to solve eq. (\ref{1}) is
equivalent to solve the null geodetics equation together with Raychaudhuri
equation since the wave fronts $S$ gives informations both on the light rays
(through its characteristics) and on the shear and on the expansion of the
light beams (through the induced geometry of the null hypersurfaces $S=const$
which can be easily studied once $S$ is given). The method here proposed,
which reduces eq. (\ref{1}) to a much simpler set of linear partial
differential equations, seems to be an effective tool to explore holography.

Now, let us consider the case in which $S$ can be a complex function:
$S=V+\sqrt{-1}W$. It is fair to say that, in this case, the WKB approximation
breaks down since, with a complex $S$, it is not true anymore that we have a
wave with a slowly varying amplitude $A$ and a rapidly varying phase
$S=V+\sqrt{-1}W$, being the amplitude multiplied by $\exp\left(  -W\right)  $
which is rapidly varying. The analysis of this case is rather interesting:
firstly, it clarifies which is the geometrical structure responsible of the
breaking of the WKB approximation. Moreover, it shed new light on some
interesting phenomena occurring in the presence of such geometrical structure.
A remark is in order here. It is rather trivial to construct special complex
solutions of eq. (\ref{1}): given a real solution of eq. (\ref{1}) $S_{0}$,
the complex function $S_{0}+\sqrt{-1}F(S_{0})$ is a complex solution of eq.
(\ref{1}) for any differentiable real function $F(x)$. Obviously, this kind of
complex solutions are uninteresting and have no physical meaning. On the
contrary, the complex solutions we will search for arise only in very special
situations and are related to geometric structures (which generalize the
Cauchy-Riemann conditions) signaling interesting physical phenomena.

When one rows of $\Theta$ vanishes, separating the real and the imaginary part
one obtains the following linear system of equations for $V$ and $W$:
\begin{align}
\left(  \varepsilon_{(0)}^{\mu}\pm\varepsilon_{(1)}^{\mu}\right)
\partial_{\mu}V  &  =0,\qquad\left(  \varepsilon_{(0)}^{\mu}\pm\varepsilon
_{(1)}^{\mu}\right)  \partial_{\mu}W=0\label{C1}\\
\varepsilon_{(3)}^{\mu}\partial_{\mu}W  &  =\mp\varepsilon_{(2)}^{\mu}%
\partial_{\mu}V,\qquad\varepsilon_{(2)}^{\mu}\partial_{\mu}W=\pm
\varepsilon_{(3)}^{\mu}\partial_{\mu}V\text{ }. \label{C2}%
\end{align}
where $\varepsilon_{(0)}^{\mu}\pm\varepsilon_{(1)}^{\mu}$ is the propagation
direction of the geodesic flows and both the real and the imaginary part of
$S$ have vanishing Lie derivative along the propagation direction. The non
trivial dependence in the transversal directions is parametrized by a sort of
generalized harmonic functions in the $(2)-(3)$ directions since eqs.
(\ref{C2}) can be thought as generalized Cauchy-Riemann conditions for $V$ and
$W$. To see this, let us consider the case in which
\begin{align*}
\varepsilon_{(2)}^{\mu}  &  =C\partial_{2}+D\partial_{3},\quad\varepsilon
_{(3)}^{\mu}=E\partial_{2}+F\partial_{3}\\
CF-DE  &  \neq0\text{.}%
\end{align*}
With the above tetrads, eqs. (\ref{C2}) read:
\begin{align}
\partial_{i}W  &  =\widetilde{T}_{i}^{j}\partial_{j}V,\quad i,j=2,3
\label{7R}\\
\qquad\widetilde{T}_{j}^{i}  &  =\frac{1}{CF-DE}\left(
\begin{array}
[c]{cc}%
EF+CD & F^{2}+C^{2}\\
-\left(  E^{2}+C^{2}\right)  & -\left(  EF+CD\right)
\end{array}
\right)  \text{,} \label{NR}%
\end{align}
thus, we see that an important role in order to have complex solutions is
played by the tensor defined in eq. (\ref{NR}) being a sort of complex
structure in the transversal directions spanned by $\varepsilon_{(2)}^{\mu}%
$\ and $\varepsilon_{(3)}^{\mu}$. In the following, some geometrical
conditions on eqs. (\ref{C2}) ensuring the possibility to have non trivial
complex solutions of HJ equation together with two interesting examples will
be analyzed.

In the case in which the two rows are proportional the equations read:
\begin{align}
\varepsilon_{(0)}^{\mu}\partial_{\mu}S+\varepsilon_{(1)}^{\mu}\partial_{\mu}S
&  =k_{C}\left(  \varepsilon_{(2)}^{\mu}\partial_{\mu}S-\sqrt{-1}%
\varepsilon_{(3)}^{\mu}\partial_{\mu}S\right) \label{5}\\
\varepsilon_{(2)}^{\mu}\partial_{\mu}S+\sqrt{-1}\varepsilon_{(3)}^{\mu
}\partial_{\mu}S  &  =k_{C}\left(  \varepsilon_{(0)}^{\mu}\partial_{\mu
}S-\varepsilon_{(1)}^{\mu}\partial_{\mu}S\right)  \label{6}%
\end{align}
$k_{C}=A+\sqrt{-1}B$ being a non zero complex function on $M$. Eqs. (\ref{5})
and (\ref{6}) can be splitted into four real equations as follows:
\begin{align}
L_{(b)}^{(a)}X_{(a)}  &  =Y_{(b)},\qquad X_{(a)}=\varepsilon_{(a)}^{\mu
}\partial_{\mu}W,\qquad Y_{(a)}=N_{(a)}^{(b)}\varepsilon_{(b)}^{\mu}%
\partial_{\mu}V,\label{C3}\\
L_{(b)}^{(a)}  &  =\left(
\begin{array}
[c]{cccc}%
0 & 0 & -B & A\\
1 & 1 & -A & -B\\
-B & B & 0 & 1\\
A & -A & -1 & 0
\end{array}
\right)  ,\quad N_{(b)}^{(a)}=\left(
\begin{array}
[c]{cccc}%
1 & 1 & -A & -B\\
0 & 0 & B & -A\\
-A & A & 1 & 0\\
-B & B & 0 & 1
\end{array}
\right) \nonumber
\end{align}
To express the imaginary part of $S$ in terms of the real part, we have to
solve the linear system (\ref{C3}) for the $X_{(a)}$. However, $\det L=0$
since only three rows are independent and the first row $\overrightarrow
{r}_{(1)}$ is a linear combination of the third $\overrightarrow{r}_{(3)}$ and
the fourth $\overrightarrow{r}_{(4)}$ rows $\overrightarrow{r}_{(1)}%
=A\overrightarrow{r}_{(3)}+B\overrightarrow{r}_{(4)}$. In order for the system
of equations (\ref{C3}) to be compatible, the following relation, which
determines the real part of $S$, must hold:
\begin{align}
Y_{(0)}  &  =AY_{(2)}+BY_{(3)}\Rightarrow\nonumber\\
&  \frac{\left(  A^{2}+B^{2}-1\right)  }{\left(  A^{2}+B^{2}+1\right)
}\varepsilon_{(1)}^{\mu}\partial_{\mu}V+\frac{2A}{\left(  A^{2}+B^{2}%
+1\right)  }\varepsilon_{(2)}^{\mu}\partial_{\mu}V+\nonumber\\
+\frac{2B}{\left(  A^{2}+B^{2}+1\right)  }\varepsilon_{(3)}^{\mu}\partial
_{\mu}V  &  =\varepsilon_{(0)}^{\mu}\partial_{\mu}V\text{.} \label{F1}%
\end{align}
By standard linear algebra's consideration, when the above equation is
satisfied, we can discard the first equation of the system (\ref{C3}) and
freely assign $\varepsilon_{(0)}^{\mu}\partial_{\mu}W$, so we will take
$\varepsilon_{(0)}^{\mu}\partial_{\mu}W=0$. Eventually, it is convenient, in
the reduced system (\ref{C3}), to express the Lie derivative of $V$ along the
timelike tetrad $\varepsilon_{(0)}^{\mu}\partial_{\mu}V$ \ in terms of the Lie
derivative along the spacelike ones $\varepsilon_{(i)}^{\mu}\partial_{\mu}V$
(with $(i)=1,2,3$) through eq. (\ref{F1}). Thus, we arrive at the following
system in which only spacelike Lie derivatives are involved:
\begin{align*}
L_{(j)}^{(i)}X_{(i)}  &  =\widetilde{N}_{(j)}^{(i)}\varepsilon_{(i)}^{\mu
}\partial_{\mu}V,\quad L_{(j)}^{(i)}=\left(
\begin{array}
[c]{ccc}%
0 & A & B\\
B & 0 & 1\\
-A & -1 & 0
\end{array}
\right)  ,\quad\det L_{(j)}^{(i)}\neq0,\\
\widetilde{N}_{(j)}^{(i)}  &  =\left(
\begin{array}
[c]{ccc}%
0 & B & A\\
\frac{2A}{A^{2}+B^{2}+1} & 1-\frac{2A^{2}}{A^{2}+B^{2}+1} & -\frac{2AB}%
{A^{2}+B^{2}+1}\\
\frac{2B}{A^{2}+B^{2}+1} & -\frac{2AB}{A^{2}+B^{2}+1} & 1-\frac{2B^{2}}%
{A^{2}+B^{2}+1}%
\end{array}
\right)  .
\end{align*}
The above equation can be formally solved for $X_{(i)}$ as follows:
\begin{equation}
X_{(i)}=\varepsilon_{(i)}^{\mu}\partial_{\mu}W=\left(  L^{-1}\circ
\widetilde{N}\right)  _{(i)}^{(j)}\varepsilon_{(j)}^{\mu}\partial_{\mu
}V\text{.} \label{pre}%
\end{equation}
Now, to have a clearer geometrical picture, let us consider an holonomic
coordinates system in which the spacelike vector fields of the tetrad basis
$\varepsilon_{(i)}^{\mu}$\ have no timelike components: $\varepsilon_{(i)}%
^{0}=0$, $\forall i=1,..,3$. This, locally, can always be achieved. In this
coordinates system, the components of the spacelike vector fields of the
tetrad basis $\varepsilon_{(j)}^{\alpha}$ (where we use the first greek
letters $\alpha$, $\beta$, $\gamma$ and so on, to denote spacelike indices)
can be seen as an invertible map between spacelike coordinates vector fields
$\partial_{\alpha}$ and the spacelike tetrad vector fields. From eq.
(\ref{pre}), we can express $\partial_{\alpha}W$ as follows:
\begin{align}
\partial_{\alpha}W  &  =T_{\alpha}^{\beta}\partial_{\beta}V\label{7}\\
T_{\alpha}^{\beta}  &  =\left(  \varepsilon_{(i)}^{\alpha}\right)
^{-1}\left(  L^{-1}\circ\widetilde{N}\right)  _{(i)}^{(j)}\varepsilon
_{(j)}^{\beta} \label{8}%
\end{align}
where $T_{\beta}^{\alpha}$ is a tensor field depending only on the metric of
the space-time and on $A$ and$\ B$. Eq. (\ref{7}) (as well as eq. (\ref{7R})
in the previous case) is related to the range of applicability of the standard
WKB approximation.

Let $V$ be a real solution of HJ equation (with suitable choices of
$k$\footnote{For example $k=1$, $B=0$ and $A=1$.} solutions of eqs.
(\ref{RE1}),\ (\ref{RE2}) and (\ref{RE3}) are solutions of eq. (\ref{F1})
too), if eq. (\ref{7}) and eq. (\ref{7R}) can be fulfilled with reasonable
boundary conditions, then eq. (\ref{7}) generates a new solution with the same
real part and an imaginary part representing a geometrical magnification or
damping factor. Thus, in these cases, the WKB approximation cannot be used
anymore because the amplitudes of the waves travelling in such regions are
multiplied by rapidly varying magnification factors of the form $\exp\left(
-W\right)  $.

The problem is to determine the geometrical properties of $T_{\beta}^{\alpha}
$ that make this possible. Clearly, not any tensor makes eq. (\ref{7}) (or eq.
(\ref{7R})) consistent: we have to impose that $\partial_{\alpha\gamma}%
^{2}W=\partial_{\gamma\alpha}^{2}W$:
\begin{equation}
\partial_{\alpha\gamma}^{2}W=\partial_{\gamma\alpha}^{2}W\Rightarrow
\partial_{\gamma}\left(  T_{\alpha}^{\beta}\partial_{\beta}V\right)
=\partial_{\alpha}\left(  T_{\gamma}^{\beta}\partial_{\beta}V\right)  \text{.}
\label{CC}%
\end{equation}
The solvability of eqs. (\ref{CC}) and the similar compatibility conditions
deduced from eq. (\ref{C2}) with reasonable boundary conditions puts strong
constraint on the topology. Let us consider a simple but interesting example,
four dimensional PP-waves (first introduced in \cite{Pe59} \cite{Pe60}):
\begin{align*}
g  &  =dt^{2}-dz^{2}+H(t-z,x,y)d\left(  t-z\right)  ^{2}-dx^{2}-dy^{2},\\
\left(  \partial_{x}^{2}+\partial_{y}^{2}\right)  H  &  =0\text{.}%
\end{align*}
In this case, eqs. (\ref{C2}), which allows the arising of a complex solution
$W$, become:
\begin{equation}
\partial_{y}W=-\partial_{x}V,\qquad\partial_{x}W=\partial_{y}V\text{ }
\label{CARI}%
\end{equation}
This Cauchy-Riemann conditions simply tell us that $V$ is a two-dimensional
harmonic function and $W$ is the harmonic function conjugated to $V$. However,
when one imposes the physical boundary condition that $V$ has to approach to a
constant in the limit $x^{2}+y^{2}\rightarrow\infty$, one only finds the
trivial solution because of the maximum principle for harmonic functions. In
the case of singular PP-waves, (which are naturally coupled to cosmic string
along the $z-$axis and/or $\gamma-$ray bursts propagating in the same
direction \cite{CV03}) the metric and the Einstein equations reduce to:
\begin{align*}
g  &  =dt^{2}-dz^{2}+H(t-z,x,y)d\left(  t-z\right)  ^{2}-dx^{2}-dy^{2},\\
\left(  \partial_{x}^{2}+\partial_{y}^{2}\right)  H  &  =\overset{N}{\sum}%
\mu_{i}\delta_{i}(x_{i}(t-z),y_{i}(t-z))
\end{align*}
where $N$ is the number of cosmic strings, $\mu_{i}$ is the energy per unit of
length and $\left(  x_{i}(t-z),y_{i}(t-z)\right)  $ is the position of the
$i-$th cosmic string in the $x-y$ plane. Even if eqs. (\ref{C2}) remain
formally the same as eqs. (\ref{CARI}), now there are singularities in the
$x-y$ plane corresponding to the positions of the cosmic strings. Thus,
harmonic functions approaching to a constant in the limit $x^{2}%
+y^{2}\rightarrow\infty$ in this case exist thanks to the singularities in the
$x-y$ plane and non trivial complex solutions of the HJ equation appear.
Another interesting example is the Reissner-Nordstrom blackhole:
\begin{equation}
g=\left(  1-\frac{2M}{r}+\frac{Q^{2}}{r^{2}}\right)  dt^{2}-\left(
1-\frac{2M}{r}+\frac{Q^{2}}{r^{2}}\right)  ^{-1}dr^{2}-r^{2}d\Omega
^{2}\text{.} \label{RN}%
\end{equation}
In the case in which the $2$ and $3$ directions are $r$ and $\varphi$ and
$\theta=\pi/2$, eqs. (\ref{C2}) are:
\begin{align}
\Gamma\partial_{r}W  &  =-\partial_{\varphi}V,\qquad\partial_{\varphi}%
W=\Gamma\partial_{r}V\Rightarrow\nonumber\\
\left[  \Gamma\partial_{r}^{2}+\frac{1}{\Gamma}\partial_{\varphi}^{2}+\left(
\partial_{r}\Gamma\right)  \partial_{r}\right]  V  &  =0,\qquad\Gamma=r\left(
1-\frac{2M}{r}+\frac{Q^{2}}{r^{2}}\right)  ^{1/2}\text{.} \label{EL}%
\end{align}
If $M>\left|  Q\right|  $, (in which case the surface gravity is non
vanishing) the operator appearing in eq. (\ref{EL}) is of mixed type: elliptic
outside the two zeros of $g_{tt}$ and hyperbolic inside. In this case, there
is no real continuous piece-wise smooth solution of eq. (\ref{EL}) approaching
to zero for $r\rightarrow\infty$ (this can be rigorously proved using standard
techniques in linear partial differential equations). However, if $M=\left|
Q\right|  $ (so that the surface gravity is vanishing), the situation changes
and real non trivial continuous (but not differentiable on the line $r=M$)
solutions appear since the operator in eq. (\ref{EL}) is elliptic degenerate
on the line $r=M$. The detailed proof of this fact is trivial but a little bit
long. It is possible to convince oneself of this by analyzing the
Majumdar-Papapetrou solution \cite{Ma47} \cite{Pa47} which represents $N$
extreme charged blackholes in equilibrium and is a BPS states in supergravity
\cite{OW99}. The metric (\ref{RN}) reads:
\begin{align*}
g  &  =U^{2}dt^{2}-U^{-2}\left(  dx^{2}+dy^{2}+dz^{2}\right) \\
\left(  \partial_{x}^{2}+\partial_{y}^{2}+\partial_{z}^{2}\right)  U  &
=\overset{N}{\sum}M_{i}\delta(x_{i},y_{i},z_{i})
\end{align*}
where $(x_{i},y_{i},z_{i})$ is the position of the $i-$th blackhole and
$M_{i}$ its mass equal to its charge. In the case in which the $2$ and $3$
directions are $x$ and $y$ and $z=z_{i}$, eqs. (\ref{C2}) look like eqs.
(\ref{CARI}), but there is a singularity in the $x-y$ plane corresponding to
the positions of the $i-$th blackhole $(x_{i},y_{i})$. Thus, non trivial
harmonic functions approaching to a constant for $x^{2}+y^{2}\rightarrow
\infty$ appear. Hence, extreme blackholes admit non trivial complex solutions
of the HJ equation. This characteristic of extreme charged blackholes is
indeed related to their BPS nature. Non trivial complex solutions appear in
the extreme case because the two dimensional operator in eq. (\ref{EL}) is
elliptic degenerate on the line $r=M$. This happens because the components
$g_{tt}$ of the metric tensor (which approaches to $1$ for large $r$) has only
one zero and does not change sign and this is a consequence of the fact that
extreme charged blackholes admit a globally defined timelike Killing vector
field\footnote{For example, in the Schwarzschild case, the Killing vector
$\partial_{t}$ is timelike outside the horizon and spacelike inside.}. The
existence of a global timelike Killing vector field is a necessary condition
in order to have a BPS states in supergravity (see the discussion of Gibbons
in \cite{OW99}). It is a well known phenomenon (see, for example, \cite{OW99}
and references therein) that BPS states in field theory often carry complex
structures similar to the ones in eq. (\ref{7}) and eq. (\ref{C2}).
Eventually, when $M<\left|  Q\right|  $, the operator appearing in eq.
(\ref{EL}) is elliptic non degenerate and, for any given Dirichlet boundary
condition, the Cauchy problem admits a unique solution. It is worth to note
that the problem to understand if a given gravitational field does or does not
represent an extreme blackhole is reduced to the analysis of Cauchy problems
for two dimensional linear partial differential equations, this can be a great
simplification in more general cases. We learn from this examples that the
solvability of eqs. (\ref{C2}) and eqs. (\ref{7}) is able to detect space-time
region with not only a strong gravitational field but also strong
electromagnetic fields and/or angular momenta.

The solvability of eqs. (\ref{C2}), (\ref{7R}) and (\ref{7}) in asymptotically
flat space-times and, in particular, its character could be related to the
surface gravity of the horizon in the general case which, unfortunately, is
very difficult to analyze. The following argument provides an explanation of
this relation in the stationary case in which the event horizon is a Killing
horizon. It is well known that if the surface gravity is non vanishing, the
Killing horizon (which can be thought as a real solution of HJ equation) is
bifurcating while, when it vanishes, there is no bifurcation (see, for
example, \cite{Ch84}). However, it should be kept in mind that, when one says
''there is no bifurcation if the surface gravity vanishes'' one usually means
that there is no \textit{real} bifurcation. In fact, when the surface gravity
approaches to zero the bifurcation phenomenon does not disappear, the new
solution simply becomes complex and eq. (\ref{CC}) tells us precisely when
such complex solutions appear. This explains why the solvability of eqs.
(\ref{7R}) and (\ref{7}), through eq. (\ref{CC}), is able to detect Killing
horizons with vanishing surface gravity, that is, extreme black holes.

For the above reasons, there is also a relation with the problem of the
\textit{Cosmic Censor} conjecture. A \textit{Cosmic Censor} should not allow
gravitational fields with generic breaking of the WKB approximation, which
characterizes extreme blackholes, while he should allow only gravitational
fields for which eqs. (\ref{C2}) and eqs. (\ref{7}) can not be satisfied, in
such a way that gravitational fields would not go beyond extremality. These
arguments shows that a deeper analysis of HJ equation with the above method
could shed light on the physical constraints needed to make \textit{Cosmic
Censor} conjecture true.

\section{Conclusions}

In this paper, a new method to deal with HJ equation for null hypersurfaces in
curved four dimensional space-times has been presented. This method, based on
a convenient form for HJ equation, simplifies the HJ equation without any
reference to the field equations for the gravitational field, so that it could
be used even in theories other then general relativity. It allows to find
explicitly the geometrical conditions on the metric which the gravitational
field has to satisfies in order to have non trivial complex solutions of the
HJ equation. This space-times, in which the WKB approximation cannot be used,
are of great importance since the origin of the breaking of the WKB scheme is
related to the presence of extreme blackholes. Thus, the analysis of the case
in which such non trivial complex solutions arise could also shed light on the
physical conditions to be imposed in order to make the \textit{Cosmic Censor}
conjecture true.

\end{document}